\documentclass[prl,aps,superscriptaddress,twocolumn,amssymb,showpacs]{revtex4-1}

\usepackage{color}
\usepackage{graphicx}
\usepackage{longtable}
\usepackage{epsfig}
\usepackage{dcolumn}
\usepackage{bm}
\usepackage{amssymb}

\begin{document}

\title{Quantum phase transitions in the Kitaev--Heisenberg model on a single hexagon}

\author{     Dorota Gotfryd }
\email{Corresponding author: dorota.gotfryd@uj.edu.pl}
\affiliation{Marian Smoluchowski Institute of Physics, Jagiellonian University,
             prof. \L{}ojasiewicza 11, PL-30348 Krak\'ow, Poland }

\author {    Andrzej M. Ole\'{s} }
\affiliation{Marian Smoluchowski Institute of Physics, Jagiellonian University,
             prof. \L{}ojasiewicza 11, PL-30348 Krak\'ow, Poland }
\affiliation{Max-Planck-Institut f\"ur Festk\"orperforschung,
             Heisenbergstrasse 1, D-70569 Stuttgart, Germany}

\date{22 June, 2014}

\begin{abstract}
We present a detailed analysis of the Kitaev--Heisenberg model
on a single hexagon. The energy spectra and spin--spin correlations
obtained using exact diagonalisation indicate quantum phase transitions
between antiferromagnetic and anisotropic spin correlations when the
Kitaev interactions increase. In cluster mean-field approach
frustrated nearest neighbor exchange stabilizes the stripe phase in
between the N\'eel phase and frustrated one which evolves towards
the Kitaev spin liquid.\\
\textit{ Published in Acta Phys. Polon. A \textbf{127}, 318 (2015).}
\end{abstract}


\maketitle

Possible realizations of quantum spin liquids is one of the most
intriguing questions in modern theory of frustrated spin systems
\cite{Nor09,Bal10,Ole12}. One of the prominent examples of spin liquid
was introduced by Kitaev \cite{Kit06}. As a unique feature of this
exactly solvable model spin--spin correlations are finite only on
nearest neighbor (NN) bonds \cite{Bas07}. Recently a lot of attention
is devoted to frustrated spin models on the honeycomb lattice, either
to $J_1$-$J_2$ Heisenberg interactions \cite{Alb11,Gan13}, or to
Kitaev-Heisenberg (KH) model \cite{Jac09,Cha10,Kim11,Sch12,Rau14}.
The latter is motivated by $A_2$IrO$_3$ iridates ($A$=Na,Li) which is
a candidate to host Kitaev model physics. For a realistic description
of these compounds, and in particular of the observed zigzag phase
\cite{Tro13}, also next nearest neighbor (NNN) and third nearest
neighbor (3NN) Heisenberg antiferromagnetic (AF) interactions
frustrating the N\'eel state are necessary \cite{Alb11,Kim11} --- these
terms are also justified by rather itinerant character of the electrons
in $A_2$IrO$_3$ \cite{Val13}. Several experiments suggest that the NNN
($J_2$) and 3NN ($J_3$) coupling constants have similar values, i.e.,
$J_2\approx J_1/2$, $J_3\approx J_2$ \cite{Kim11}.

The purpose of this paper is to investigate the evolution of spin--spin
correlations on a single hexagon when interactions change from
AF Heisenberg to highly frustrated ferromagnetic (FM) Kitaev ones.
This evolution is modified when a cluster mean-field (MF) approach is
applied, similar to the one used before for the $J_1$-$J_2$-$J_3$
model \cite{Alb11} and Kugel-Khomskii model \cite{Brz12}.

The KH Hamiltonian has the form \cite{Kim11},
\begin{eqnarray}
H&\equiv&
-2J\alpha\sum_{\langle ij\rangle\parallel\gamma} S_i^\gamma S_j^\gamma
+ J(1-\alpha)\Big\{\sum_{\langle ij\rangle}   \vec{S}_i\cdot\vec{S}_j
\nonumber \\
&+&J_2\sum_{\langle\langle ij \rangle\rangle}\vec{S}_i\cdot\vec{S}_j
 + J_3\sum_{\langle\langle\langle ij\rangle\rangle\rangle}
\vec{S}_i\cdot\vec{S}_j\Big\}\,.
\label{mm}
\end{eqnarray}
In the first Kitaev term, bond-dependent Ising-like interactions are
selected by $\gamma\in\{x,y,z\}$ depending on the bond direction.
The parameter $\alpha$ interpolates between Heisenberg ($\alpha=0$)
and Kitaev ($\alpha=1$) interactions. We set the energy unit $J=1$,
and we take equal NNN ($J_2$) and 3NN ($J_3$) interactions, i.e.,
$J_2=J_3=J_1/2$ and $J_1\equiv(1-\alpha)J$. Following the
{\it ab initio\/} calculations \cite{Val13}, we select the AF NN
Heisenberg terms and FM Kitaev ones. Note that already at small
$\alpha>0$ spin interactions are anisotropic, and classically N\'eel or
resonating valence bond (RVB) phase is destroyed at $\alpha=1/3$ when
some NN interactions switch from AF to FM. Here we investigate the
more challenging quantum case.

We performed exact diagonalisation (at $T=0$) and investigated the
energy spectra and spin correlations between NN, NNN, and 3NN
spins at sites $\{i,j\}$,
\begin{equation}
S(i,j)= \langle\vec{S}_{i}\cdot\vec{S}_{j}\rangle=\frac{1}{d}
\sum_{k=1}^d\langle\Phi_k|\vec{S}_i\cdot\vec{S}_j|\Phi_k\rangle,
\label{ss}
\end{equation}
where $\{|\Phi_k\rangle\}$ are individual degenerate states in the
ground state manifold, and $k=1,\dots,d$. In addition, we investigate
below {\it partial\/} spin correlations which reflect the anisotropic
character of spin interactions,
\begin{equation}
S^{\gamma}(i,j)= \langle S^{\gamma}_iS^{\gamma}_j\rangle=\frac{1}{d}
\sum_{k=1}^d\langle\Phi_k|S^{\gamma}_iS^{\gamma}_j|\Phi_k\rangle.
\label{pass}
\end{equation}
For a free hexagon, no order may occur and $\langle S_i^z\rangle\equiv 0$.

In the quest of quantum phase transitions (QPTs) several trails have
been revealed. First clue appears to be change of the ground state of
the Hamiltonian operator which defines the QPT. Second track signalling
directly the transition is the variation of spin-spin correlations ---
either the change of sign, or discontinuities which are fingerprints of
QPTs. Finally, extremal values of the ground state energy $E_0$ might
also indicate a transition \cite{Cha10}.

\begin{figure}[t!]
\includegraphics[width=7.2cm]{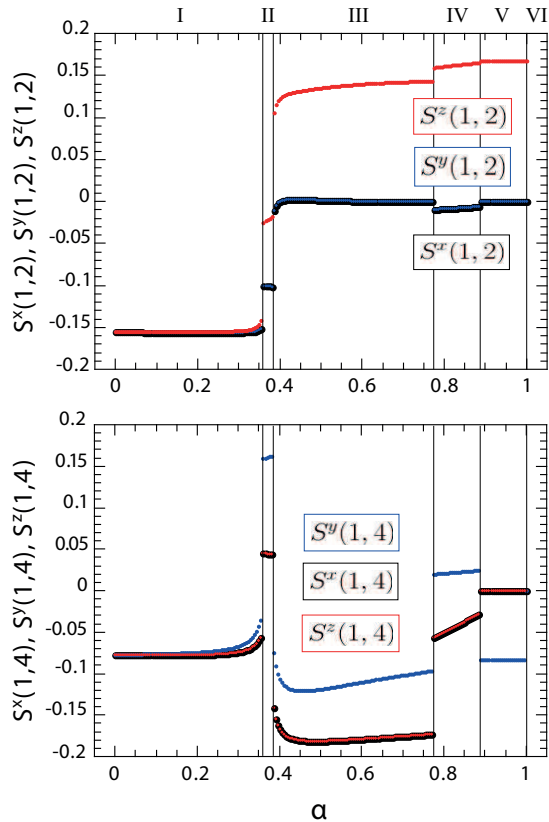}
\vskip -0.2cm
\caption{
Spin--spin correlations as obtained for KH model (1) at increasing
$\alpha$, with $-2JS_1^zS_2^z$ interaction at $\alpha=1$:
top (a) --- NN $\langle S^{\gamma}_1S^{\gamma}_2\rangle$, and
bottom (b) --- 3NN $\langle S^{\gamma}_1S^{\gamma}_4\rangle$
(this bond is parallel to $\langle S^y_2S^y_3\rangle$ at $\alpha=1$).
QPTs occur at vertical lines and phases are labeled from I to VI.}
\label{korelacje}
\end{figure}

\begin{figure}[t!]
\includegraphics[width=8.8cm]{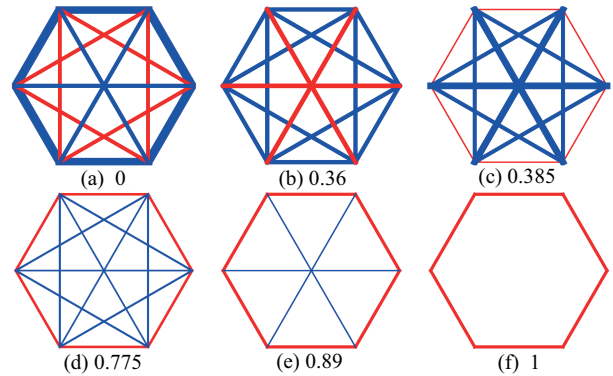}
\vskip -0.2cm
\caption{Spin--spin correlation functions for the Hamiltonian
$(\ref{mm})$ for selected values of $\alpha$ (below each panel).
The sign (AF or FM) is indicated by line color: blue --- AF
correlation, red --- FM correlation;
the line width is proportional to its absolute value.
Parameters: $J_2=J_3=0.5J_1$.}
\label{hexagony}
\end{figure}

Spin--spin correlations change in a discontinuous way at some values of
$\alpha$ which indicate QPTs. Here we show only the correlations for NN
and for 3NN which are sufficient to conclude about the QPTs when
$\alpha$ increases, see Fig. \ref{korelacje}. First, for
$\alpha\in[0, 0.355)$ (phase I), the NN correlations are almost
independent of $\gamma$, i.e., $S^{\gamma}(1,2)\simeq S(1,2)/3$,
and one finds a RVB phase which weakens above $\alpha\simeq 0.3$.
At $\alpha\simeq 0.355$ the first QPT occurs, see
Figs. \ref{korelacje}(a) and \ref{korelacje}(b), and both $S(1,3)$ and
$S(1,4)$ change signs, cf. Figs. \ref{hexagony}(a) and
\ref{hexagony}(b). Two nondegenerate states cross at the QPT
and the derivative of $E_0$ changes (Table I). As in spin-orbital
systems \cite{Brz12}, phase II is driven here by $J_2$ and $J_3$ while
$J_1$ changes sign. It has FM (AF) NNN (3NN) correlations,
see Fig. \ref{hexagony}(b), and we suggest that it is a precursor of
the zigzag phase found in this range of parameters \cite{Kim11,Tro13}.

A second QPT occurs at $\alpha\simeq 0.385$, where two nondegenerate
ground states intersect and $E_0$ is maximal. Here both spin--spin
correlations $S(1,2)$ and $S(1,4)$ change signs.
Already at $\alpha=0.355$ we observe that $S^z(1,2)$ separates from
$S^x(1,2)=S^y(1,2)$, and $S^y(1,4)$ separates from $S^x(1,4)=S^z(1,4)$,
and this persists up to $\alpha=1$, see Figs. \ref{korelacje}(a) and
\ref{korelacje}(b).

Further discontinuities arise for all $S^{\gamma}(1,2)$ at
$\alpha\simeq 0.770$, but in their sum $S(1,2)$ they nearly cancel
one another and the discontinuity of $S(1,2)$ almost vanishes. At this
QPT a singlet and a triplet cross. Notably, the correlation functions
do not change signs at this QPT, see Figs. \ref{hexagony}(c) and
\ref{hexagony}(d). We observe that in phase IV NN FM correlations grow
stronger while NNN and 3NN correlations (both AF) weaken.

At $\alpha\simeq 0.890$ the triplet state crosses with another singlet
ground state, indicating a QPT to a distinct spin disordered phase V,
stable for $\alpha\in[0.89, 1)$. All spin--spin correlations are
discontinuous at the transition (Table I) and all NNN ones vanish, see
Fig. \ref{hexagony}(e), while $S^y(1,4)$ is small and finite
\cite{Tro11}, see Fig. \ref{korelacje}(b). The gap between the ground
state and triplet excited state first grows and then start to shrink
with increasing $\alpha$ until both states merge at $\alpha=1$, where
one finds FM spin correlations for NN only, see Fig. \ref{hexagony}(f).
The only finite spin--spin correlation at $\alpha=1$ happens to be
$S^z(1,2)$, see Fig. \ref{korelacje}(a).

\begin{table}[b!]
\caption{Discontinuities in spin--spin correlations $S(1,n)$ and the
feature of the ground state energy $E_0$ (if any) at five QPTs which
occur at $\alpha_c$. At the first three QPTs spin correlations change
sign (sign) between the ground states with degeneracies $d_<$ and $d_>$
for $\alpha<\alpha_c$ and $\alpha>\alpha_c$, respectively. }
\begin{ruledtabular}
\begin{tabular}{cccccc}
 $\alpha_c$    & $S(1,n)$ &  sign & $d_<$ & $d_>$ & feature of $E_0$ \cr
\colrule
 $\sim 0.355$  & $S(1,3)$ & $+/-$ & 1 & 1 &  slope change \cr
               & $S(1,4)$ & $-/+$ &   &   &               \cr
 $\sim 0.385$  & $S(1,2)$ & $-/+$ & 1 & 1 &    maximum    \cr
               & $S(1,4)$ & $+/-$ &   &   &               \cr
 $\sim 0.770$  & $S(1,4)$ & $+/-$ & 1 & 3 &  slope change \cr
 $\sim 0.890$  & $S(1,n)$ &$\dots$& 3 & 1 &    $\dots$    \cr
      $1.0$    & $S(1,4)$ & $-/0$ & 1 & 4 &    $\dots$    \cr
\end{tabular}
\end{ruledtabular}
\label{tabela2}
\end{table}

For $\alpha=1$ the ground state degeneracy is $d=4$; it is lifted when
minute Heisenberg interaction is added at $\alpha<1$, in analogy to the
2D compass model, where Heisenberg terms remove high degeneracy of the
ground state \cite{Tro12}. In contrast, however, the ground state does
not change and the Kitaev spin liquid survives here in the range of
$\alpha\in[0.89,1)$, with additional 3NN correlations.

Special attention has to be paid to $S^{y}(1,4)$, with its sign being
different from that of $S^x(1,4)=S^z(1,4)$ when $\alpha\in[0.355,1)$.
This function has a discontinuity at each QPT, see Table I.
It concerns the bond $\langle 14\rangle$ which is parallel to the NN
bond $\langle 23\rangle$ with $S^y_2S^y_3$ interaction in the Kitaev
limit, so we see that the Kitaev part induces 3NN correlations for the
same component $\gamma$ which is active along the NN bonds parallel to
it. Partial NNN spin correlations also separate at $\alpha=0.355$ but
drop to zero when spins get disordered at $\alpha=0.890$.

\begin{figure}[t!]
\begin{center}
\includegraphics[width=7.7cm]{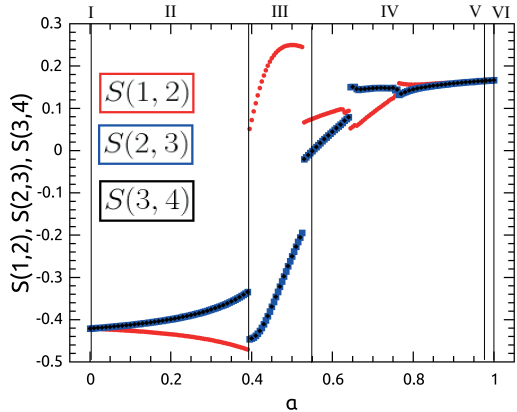}
\end{center}
\vskip -.2cm
\caption{
Spin--spin NN correlations obtained for converged MF calculations
for $0\leq\alpha\leq 1$: $S(1,2)$, $S(2,3)$, $S(3,4)$.
QPTs occur at vertical lines and phases are labeled from I to VI.
Parameters: $J_2=J_3=0$.}
\label{fig:mf}
\end{figure}

Previous studies within the cluster MF \cite{Alb11,Brz12} inspired us
to consider the hexagon with only NN Heisenberg $J_1$ and Kitaev
$J_K\equiv -2J\alpha$ terms. We embedded the hexagon by the MF terms,
replacing spins along outer NN bonds with the order parameters,
$s_i^z\equiv\langle S_i^z\rangle$. They were selected using either
N\'eel or stripe ansatz and calculated self-consistently. For
$\alpha\in(0,0.39]$ (phase II) the SU(2) symmetry is broken and
$\{s_i^z\}$ and $\{S(i,j)\}$ follow N\'eel AF order (phase I) which
extends up to $\alpha=0.395$ due to quantum fluctuations, see Fig.
\ref{fig:mf}. Near the QPT at $\alpha=0.390$ one finds robust N\'eel
order with positive/negative values of $|s_i^z|\simeq 0.4172$ at
odd/even site $i$ of the hexagon.

For $\alpha\leq 0.36$ the stripy ansatz gave $s_i^z=0$, while
spin--spin correlations are constant and RVB-like. At $\alpha=0.365$
the symmetry is broken ($s_i^z\neq 0$), but the NN correlations do not
follow the stripy pattern yet. We obtained the stripy phase for
$\alpha\in[0.395,0.55]$ (phase III), with FM (AF) spin--spin
correlations $S(1,2)=S(4,5)$ (otherwise), see Fig. \ref{fig:mf}.
Unlike in N\'eel phase, here one finds two distinct values of the order
parameters $\{|s_i^z|\}$, e.g. $s_i^z\simeq 0.3795$ ($-0.2675$) for
$i=1,2,4,5$ ($i=3,6$) at $\alpha=0.395$, as the sites are nonequivalent
and the latter ones are exposed to enhanced quantum fluctuations within
the hexagon. These fluctuations disappear at $\alpha=0.5$, in agreement
with the mapping on the FM Heisenberg model \cite{Cha10}. Unfortunately,
we could not obtain converged results for $\alpha\in[0.5,0.525)$. The
region of (stripe) phase III agrees partly with that obtained for a
larger cluster of $N=24$ sites, $\alpha\in[0.4,0.8]$ \cite{Cha10}.
We thus conclude that the stripy order is subtle and hard to stabilize
on a single hexagon.

For $\alpha\in(0.555,0.98)$ (phase IV) the symmetry remains broken but
the stripe phase is destroyed here by Kitaev terms and all NN $S(i,j)$
are weakly FM and anisotropic, see Fig. \ref{fig:mf}. At $\alpha=0.98$
one finds a QPT to disordered spin liquid with $d=3$ (phase V). It is
similar to phase IV of a free hexagon (see Table I). The last QPT is
found at the Kitaev limit $\alpha=1$ itself, where we find again $d=4$.

Summarizing, we conclude that increasing Kitaev interactions cause
spin--spin correlations $S_i^\gamma S_j^\gamma$ to separate.
This phenomenon is generic and occurs both for a free hexagon and in
MF shortly after one NN interaction $S_i^\gamma S_j^\gamma$ changes
sign. Unless Kitaev terms dominate, investigation of possible
long-range order requires cluster MF or even more sophisticated methods.
The Kitaev spin liquid phase extends to $\alpha<1$ also in the MF
approach, but 3NN spin correlations are induced in this regime.

\acknowledgments

We thank Ji\v{r}i Chaloupka for insightful discussions
and acknowledge support by the Polish National Science
Center (NCN) under Project No. 2012/04/A/ST3/00331.

\end{document}